\documentclass[preprint,12pt]{elsarticle}

\usepackage{amssymb}
\usepackage{color, graphicx}

\journal{Planetary and Space Science}

\def\lex{{\raise2pt\hbox{$<$}}\llap{\lower2pt\hbox{{$\sim$}}}}
\def\b#1{\mbox{\boldmath$#1$}}

\begin{document}

\begin{frontmatter}

%% Title, authors and addresses

%% use the tnoteref command within \title for footnotes;
%% use the tnotetext command for theassociated footnote;
%% use the fnref command within \author or \address for footnotes;
%% use the fntext command for theassociated footnote;
%% use the corref command within \author for corresponding author footnotes;
%% use the cortext command for theassociated footnote;
%% use the ead command for the email address,
%% and the form \ead[url] for the home page:
%% \title{Title\tnoteref{label1}}
%% \tnotetext[label1]{}
%% \author{Name\corref{cor1}\fnref{label2}}
%% \ead{email address}
%% \ead[url]{home page}
%% \fntext[label2]{}
%% \cortext[cor1]{}
%% \address{Address\fnref{label3}}
%% \fntext[label3]{}

\title{The daily processing of asteroid observations by Gaia}

%% use optional labels to link authors explicitly to addresses:
%% \author[label1,label2]{}
%% \address[label1]{}
%% \address[label2]{}

\author{
Paolo Tanga$^{a}$,  
Fran\c{c}ois Mignard$^{a}$,
Aldo Dell'Oro$^{g}$,
Karri Muinonen$^{d,b}$,
Thierry Pauwels$^i$, 
William Thuillot$^{e}$
J\'{e}r\^{o}me Berthier$^{e}$,
Alberto Cellino$^f$, 
Daniel Hestroffer$^{e}$,
Jean-Marc Petit$^h$,
Benoit Carry$^{e}$,
Pedro David$^{e}$,
Marco Delbo'$^a$,
Grigori Fedorets$^{d}$,
Laurent Galluccio$^a$,
Mikael Granvik$^{d,b}$, 
Christophe Ordenovic$^a$,
Hanna Pentik\"ainen$^{d}$\fnref{info1} 
}

% \thanksref{info}}

 \address{$^a$Laboratoire Lagrange, Universit\'e C\^ote d'Azur, Observatoire de la C\^ote d'Azur, CNRS, Blvd de l'Observatoire, CS 34229, 06304 Nice cedex 4, France}
 \address{$^g$INAF - Osservatorio Astrofisico di Arcetri, Largo E. Fermi 5, 50125 Firenze, Italy}
   \address{$^d$Department of Physics, Gustaf H\"allstr\"omin katu 2a, FI-00014 University of Helsinki, Finland}
  \address{$^b$Finnish Geospatial Research Institute, Geodeetinrinne 2, 
    FI-02430 Masala, Finland}
  \address{$^e$Observatoire de Paris,
IMCCE, Institut de m\'ecanique c\'eleste et de calcul des \'eph\'em\'erides, Unit\'e Mixte de Recherche UMR-CNRS 8028, 77 avenue Denfert-Rochereau, F-75014 Paris, France}
  \address{$^f$INAF - Osservatorio Astrofisico di Torino, Strada Osservatorio 20, 10025 Pino Torinese, Italy}
  \address{$^i$Observatoire Royal de Belgique, Avenue Circulaire 3, B-1180 Bruxelles,
Belgique}
  \address{$^h$Observatoire de Besan\c{c}on, UMR CNRS 6213, 41 bis avenue de l'Observatoire, F-25000 Besan\c{c}on, France}

  \fntext[info1]{Corresponding author is Paolo Tanga;  Email: Paolo.Tanga@oca.eu}

\begin{abstract}
%% Text of abstract
The Gaia mission started its regular observing program in the summer of 2014, and since then it is regularly obtaining observations of asteroids. This paper draws the outline of the data processing for Solar System objects, and in particular on the daily ``short-term'' processing, from the on-board data acquisition to the ground-based processing. We illustrate the tools developed to compute predictions of asteroid observations, we discuss the procedures implemented by the daily processing, and we illustrate some tests and validations of the processing of the asteroid observations. Our findings are overall consistent with the expectations
concerning the performances of Gaia and the effectiveness of the developed
software for data reduction.
\end{abstract}

\begin{keyword}
  Gaia mission; astrometry; asteroids
%% keywords here, in the form: keyword \sep keyword

%% PACS codes here, in the form: \PACS code \sep code

%% MSC codes here, in the form: \MSC code \sep code
%% or \MSC[2008] code \sep code (2000 is the default)

\end{keyword}

\end{frontmatter}

%% \linenumbers

%% main text

\section{Introduction: Gaia and Solar System objects}
\label{sec:intro}

The European mission Gaia observes the whole sky from the Lagrangian point L2, where the required thermal stability is guaranteed (details and capabilities are described in detail by Prusti 2013, De Brujine et al. 2012, and references therein). The satellite operates in continuous scanning mode, its spin being of 6~h. Two lines of sight separated on the scanning plane by 106.5$^\circ$ (the basic angle), are simultaneously imaging the sky on the same focal plane. This feature, reducing the measurements of large angular separations to small distances on the focal plane, is the essential principle allowing Gaia to have a homogeneous all-sky astrometric accuracy, without zonal errors. The slow change in the orientation of the scanning plane, steered by a 62.97-days precession and by the 1-year revolution around the Sun, determines a rather homogenous coverage of the sky resulting, over 5 years of nominal mission duration, in 80-100 observations for an average direction, slightly less on the ecliptic (60-70). 

The images formed on the focal plane, consisting of a large Giga-pixel array of 106 CCDs, are electronically tracked on the CCD itself by a displacement of the charge (Time Delay Integration mode, TDI) at the same pace as the image drifts due to the spacecraft rotation. 

The CCDs are organized in the order of crossing by the drifting images. First, there are two CCD strips devoted to source detection (one for each of the two lines of sight); they constitute the instrument called Sky Mapper (SM). Then, 9 strips of astrometric CCDs follow (Astrometric Field, AF). Next, other CCD strips are devoted to low resolution spectro-photometry (red and blue photometer, RP/BP) and high resolution spectroscopy (Radial Velocity Spectrometer, RVS). RVS is not considered for asteroid studies, due to its narrow range of wavelength. 

Each source that enters the field of view of Gaia will produce a signal on one SM CCD. If bright enough (V$<$20.7 is the current threshold) and nearly point-like (about $<$600 mas diameter) its position is then recorded by the on-board Video Processing Unit (VPU). The VPU automatically assigns a ``window'' around each object detected by the SM, and propagates these windows to the other CCDs in the direction of the image drift. Only these very small windows (the smallest, but more common ones spanning 6 pixels only) are transmitted to Earth, in such a way that the telemetry does not exceed the possible downlink rate. Due to this windowing strategy, two point-like sources separated by more than $\sim$300 mas (6 pixels) are detected as two different images and processed separately.

Due to its orbital motion, a Solar System Object (SSO) may leave the transmitted window before arriving at the last CCD. As a consequence, each “observation” consists of a maximum of 10 positions (AF and SM instruments), distributed over 50 seconds (the duration of a transit). 

One should note that the cut--off at magnitude V=20.7 is not dictated by a threshold on the minimum, acceptable signal-to-noise ratio, as at this brightness level very accurate astrometry can still be obtained. Rather, the limit is imposed by constraints on the data downlink rate, especially in the densest areas of the Milky Way.

All source identifications and further processing are done on the ground and are part of the activities of the Data Processing and Analysis Consortium (DPAC). Also, DPAC is in charge of running the Astrometric Global Iterative Solution (AGIS), a highly optimized software system that looks for the best-fitting self-consistent attitude and astrometric solution on the sphere, taking into account all measurements and instrument calibration parameters. The astrometry based on the best AGIS result is used for the preparation of each intermediate release.

Starting from 2006, DPAC of Gaia was charged by ESA for implementing the data processing pipelines that will deliver the first-level analysis of Gaia observations. The Gaia outcome –- in fact –- will consist not only of the individual measurements, but also of calibrated data (fluxes, positions, spectra), global statistics, and the results of the exploration of the bulk properties of the sources (classification, distributions etc). 

In this context, the Coordination Unit 4 (CU4) has the task of performing the analysis of objects deserving a specific treatment, namely multiple stellar systems, exoplanets (PI: D. Pourbaix, Brussels Univ.), Solar System objects (PI: P. Tanga) and extended sources (Ch. Ducourant, Obs. Bordeaux, France). 

All the software produced within DPAC runs at Data Processing Centers; the Data Processing Center CNES (DPCC) in Toulouse, France, is in charge of Solar System data, among others. Essentially, the processing will proceed blindly for the whole DPAC community. This approach, along with the absence of any proprietary period, ensures that the data products of Gaia will be available to the whole scientific community (including the DPAC scientists) at the same time, as established by the ESA-DPAC agreement.

Gaia will obtain during its 5-year operation $\sim$70 observations per object, on average, for about 350~000 asteroids.

We recall here that the scientific community was made aware of unexpected technical difficulties (in particular, the presence of stray-light) discovered during commissioning. Recent studies of these issues reveal that they will not affect the revolutionary potential of Gaia, with a very modest degradation in the expected performance (De Bruijne, 2015).

The DPAC CU4 has implemented two pipelines for Solar System processing (Tanga et al. 2007, Mignard et al. 2007):
\begin{itemize}[-]
\item SSO-ST: the ``Solar System short-term processing'' is devoted to alert a ground-based network (Gaia-FUN-SSO, steered by IMCCE, Observatoire de Paris) in case a new asteroid is discovered. This pipeline will be running daily at DPCC (CNES in Toulouse) and is also used to verify and monitor the quality of the data received by Gaia.
\item SSO-LT: the ``Solar System long-term processing'' will run for the data releases and perform a more sophisticated data reduction with the best possible astrometric solution and the advanced instrument calibrations. Also, it will eventually perform the global data reduction by executing tasks that require the largest possible set of observations.  
\end{itemize}

The fist intermediate data release is planned for mid-2016, and is expected to  provide data for not less than 90$\%$ of the sources observed by Gaia.

The SSO-ST chain is currently running at CNES for the validation of the data processing. This implies that the observations being processed are concerning --- for the time being --- known asteroids. This situation offers several opportunities for validating the performances of Gaia on asteroid detection, and for tuning the SSO-ST pipeline.

The goal of this paper is to illustrate the main processing steps of SSO-ST. First of all, we explain the approach and the performance of the software that we developed for predicting the observations by Gaia (Sect. \ref{S:predict}), an essential validation tool. Then we review the SSO-ST pipeline step by step, starting from asteroid identification (Sect. \ref{S:ident}). The processing continues with the measurement of asteroid positions on the focal plane (Sect. \ref{S:signal}) and the subsequent coordinate transformations (Sect. \ref{S:astrometry}) toward the sky reference. The observations of a target are grouped together and an orbital solution is determined. As the goal of SSO-ST is to provide a first, approximate orbit for the recovery of new objects from the ground, a statistical approach is adopted (Sect. \ref{S:inversion}). We conclude by describing the ground--based follow--up activities (Sect. \ref{S:followup}).

\section{Prediction of Solar System observations}
\label{S:predict}

To define more precisely the quality of the observations with respect to expectations, we can exploit the simulations produced by a software developed by F.~Mignard and P.~Tanga at Laboratoire Lagrange (OCA, Nice). This unique tool exploits the very stable scanning law and the full orbital data set from the Minor Planet Center to predict when and how often a source will be seen by Gaia. The accuracy of the predictions and crossing times, compared to real Gaia data, are excellent, so that reliable statistics can be built. 

The transit predictor has been developed within the CU4/SSO in order to be able to compute in advance the observations of Solar System objects to be seen by Gaia during its operations. 

The software is an outgrowth of a detection simulator used and maintained over the years, since the very preliminary studies on Gaia, based on similar overall principles, but aiming at accurate individual transit data instead of an overall statistical relevance. In the earlier phase some approximations were acceptable (such as the 2-body Keplerian motion). The same liberty was used for the Gaia orbit about L2, in absence of other constraints before launch.

Moving to a predictor of what actually happens during the real mission implied a more rigourous modelling of the mission environment and of the dynamical modelling of the planetary motion. With the predictor the use of an exact Gaia scanning law is mandatory to reproduce the actual pointing of each FOV. Similarly, the Gaia orbit should be as close as possible to the true path of Gaia on its Lissajous orbit. Finally, the orbital elements of the asteroids must be taken to full accuracy at a reference epoch and then the position and velocity must be propagated with planetary perturbations and numerical integration instead of the simplified 2-body problem.

The program essentially solves for any $i$th asteroid and for each Field Of View $F$ over an interval of time~$[T_b, T_e]$, the following equation in $t$

\begin{equation}\label{main}
\mathbf{G}_F(t) = \mathbf{U}_i(t)
\end{equation}

where $\mathbf{U}_i(t)$ is the unit vector of the asteroid proper direction at time $t$ and  $\mathbf{G}_F(t)$ stands for the pointing direction of Gaia FOV $F$. The left-hand-side is the Gaia attitude model, here the Nominal Scanning Law, while the right-hand-side resulted from the integration of the planetary motion.

The adopted position of Gaia is provided by the Gaia Mission Operation Center, as reconstructed from telemetry over the past epochs. Afterwards, the targeted orbit is used, which is always within 7000 km of the actual orbit.

Over a certain interval of time the program finds all the roots $t_1, t_2, \cdots, t_{r_k} $  of  Eq.~\ref{main}. The solutions are found with an iterative process to locate a first approximation within a spin period and then accurately compute the solution with a Newton-Raphson method. The software has been strongly optimized for speed and allows to run a prediction for $\sim 500,000$ asteroids over 5 years in less than one hour of CPU time on a desk-computer, with output files reaching 1GB. 

The positions and velocity of the asteroids are computed by a numerical integration from the osculating epoch, using gravitational perturbations from the 8 planets (Mercury to Neptune) with the main component of the relativistic contribution.

The solar term with relativistic effect is computed as,

\begin{equation}\label{dynamic}
    \frac{d\mathbf{v}}{dt} = - \frac{GM_\odot \,\mathbf{r}}{r^3} + \frac{GM_\odot}{c^2r^3} \left( 4GM_\odot \frac{\mathbf{r}}{r}- v^2\mathbf{r} +4(\mathbf{r}\cdot\mathbf{v}) \mathbf{v} \right)
 \end{equation}

with $\mathbf{r} = \mathbf{r}_p - \mathbf{r}_\odot$ for the heliocentric position vector of the asteroid.

The planetary perturbations are given by,

\begin{equation}\label{planetary}
   \sum_k GM_k\left[\frac{\mathbf{r}_k-\mathbf{r}}{|\mathbf{r}_k-\mathbf{r}|^3}- \frac{\mathbf{r}_k}{r_k^3}\right]
\end{equation}
where $\mathbf{r}_k$ is the heliocentric position vector of the $k$th planet. Solar System ephemerides are taken from INPOP10e expressed in the barycentric frame with ICRF orientation and using TCB as independent variable.

There are at least three sources of uncertainty to consider:
\begin{itemize}[-]
  \item The computational accuracy.
  \item The position of the asteroid on the sky, including its orbit uncertainty.
  \item The position on the Gaia Focal Plane Assembly and the associated transit times.
\end{itemize}

This refers to the numerical solution of the transit equation (Eq.~\ref{main})  and to the numerical integration of the planetary dynamical equations, assuming all other parameters are exactly known. Convergence to the transit time is achieved to better than 1 ms. Other computations have the accuracy permitted by the numerical representation of numbers, which, apart from the epoch, is not a source of concern. The numerical integration of the asteroid motion over an interval of time that could reach 5 years is also compatible with a sub-mas astrometric accuracy. This is fully sufficient for the purpose of the transit predictor.

The quality of the prediction of the gaiacentric position is primarily determined by the knowledge of the osculating elements, rather than by the dynamical model, and by the predicted Gaia orbit. It is not easy to figure out how good the osculating elements are for every asteroid. As a rule of thumb for numbered asteroids (those with an IAU definitive number) the proper position is generally better than 0.5 arcsec and often better than 0.2 arcsec. 

The uncertainty stemming from the Gaia orbit itself can be easily estimated. As mentioned above, there is a requirement that the actual Gaia orbit is always within 7000 km of the predicted orbit, dictated by the optimization of the scanning law for the relativistic experiments of light deflection. Assuming an asteroid at 2.5~au, this uncertainty in the Gaia barycentric position translates into 4 arcsec for the asteroid Gaiacentric direction.

\begin{figure}[h]
  \begin{center}
    \includegraphics[width=0.8\columnwidth]{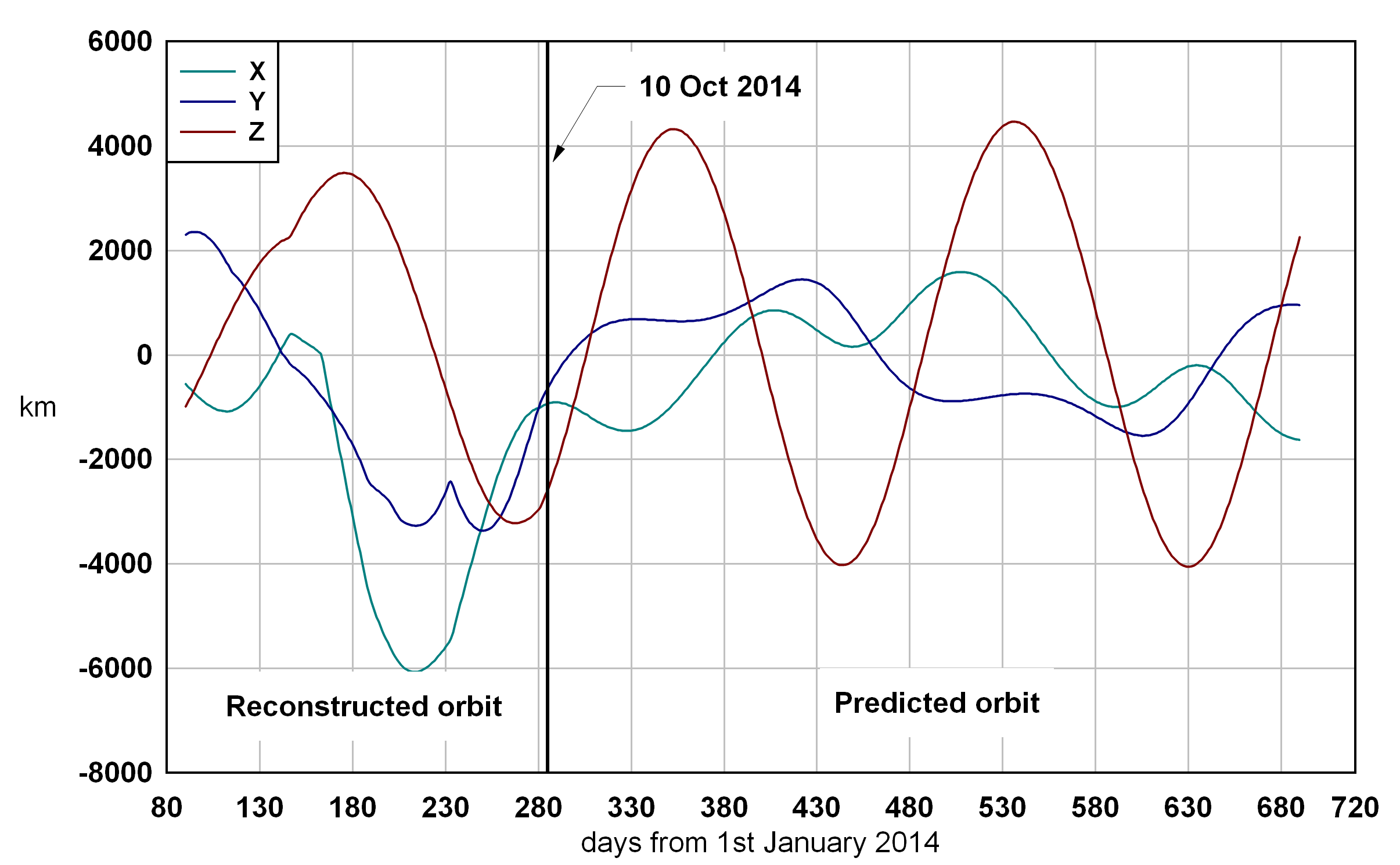}
  \end{center}
  \vspace{-5mm}
\caption{Difference between the MOC orbit by the Gaia Mission Operation Center of ESA (MOC) provided on 14 October 2014, compared to the first post-launch predicted orbit of 30 December 2013. The colors represents the differences along the (X,Y,Z) cartesian coordinates. Until 10 October 2014 the comparison is between the reconstructed and the predicted orbits, while after this is between the two predicted orbits. The large difference (still within the requirements) is due to the origin of the predicted orbit, starting 80 days before the origin of the plot. On the short term (a few days after the prediction) the situation is better and the divergence builds up gradually.}
\label{Fig:orbits}
\end{figure}

Comparisons of successive releases of the Gaia orbits indicate that the 7000 km requirement is met as shown in Fig.~\ref{Fig:orbits}, where the first predicted orbit is compared to the actual orbit until 10 October 2014 and to the most recent predicted orbit afterwards. One may assume that in 2015 the difference between the actual Gaia orbit and the one used in this version of the software will be similar giving then a maximum uncertainty as large as 4 arcsec in the predicted Gaiacentric direction of the asteroids, but only 1.5 arcsec RMS. For the numbered asteroids with good orbital elements, the uncertainty in the Gaia orbit could be the largest single factor in the overall uncertainty of the proper direction.

The main source of uncertainty here is the use of the Nominal Scanning Law of Gaia to compute the satellite attitude instead of the true attitude (not known in fact for the future!). Comparisons between daily attitude solution to the nominal scanning law have shown that the actual attitude does not differ from the targeted scanning by more than 30 arcsec and that it is very often less than 15 arcsec.

This gives an error $\sim$0.5~s in the crossing time and 100-200 pixels in the direction perpendicular to the scan. Comparisons to true Gaia observations processed by the Initial Data Treatment (IDT, implementing the first, approximate astrometric reduction) show that the difference is almost always less than 100 pixels across-scan (corresponding to 0.4~s). Limited test trials with the daily attitude have reduced this difference to few pixels and 0.03~s in the transit time, thus producing a additional validation of the implementation.

During the ACM2014 meeting, the detection of the asteroid (4997) Ksana has been presented as a validation of the capabilities of Gaia and of the accuracy of the predictor. The sky position of this specific asteroid was within 2 arcsec from the prediction. Further tests with a larger population of asteroids show that -- when a multi-opposition, good quality orbit is available, the discrepancy is more of the order of $\sim$200 mas.
\begin{figure}
	\centering
		\includegraphics[width=0.9\columnwidth]{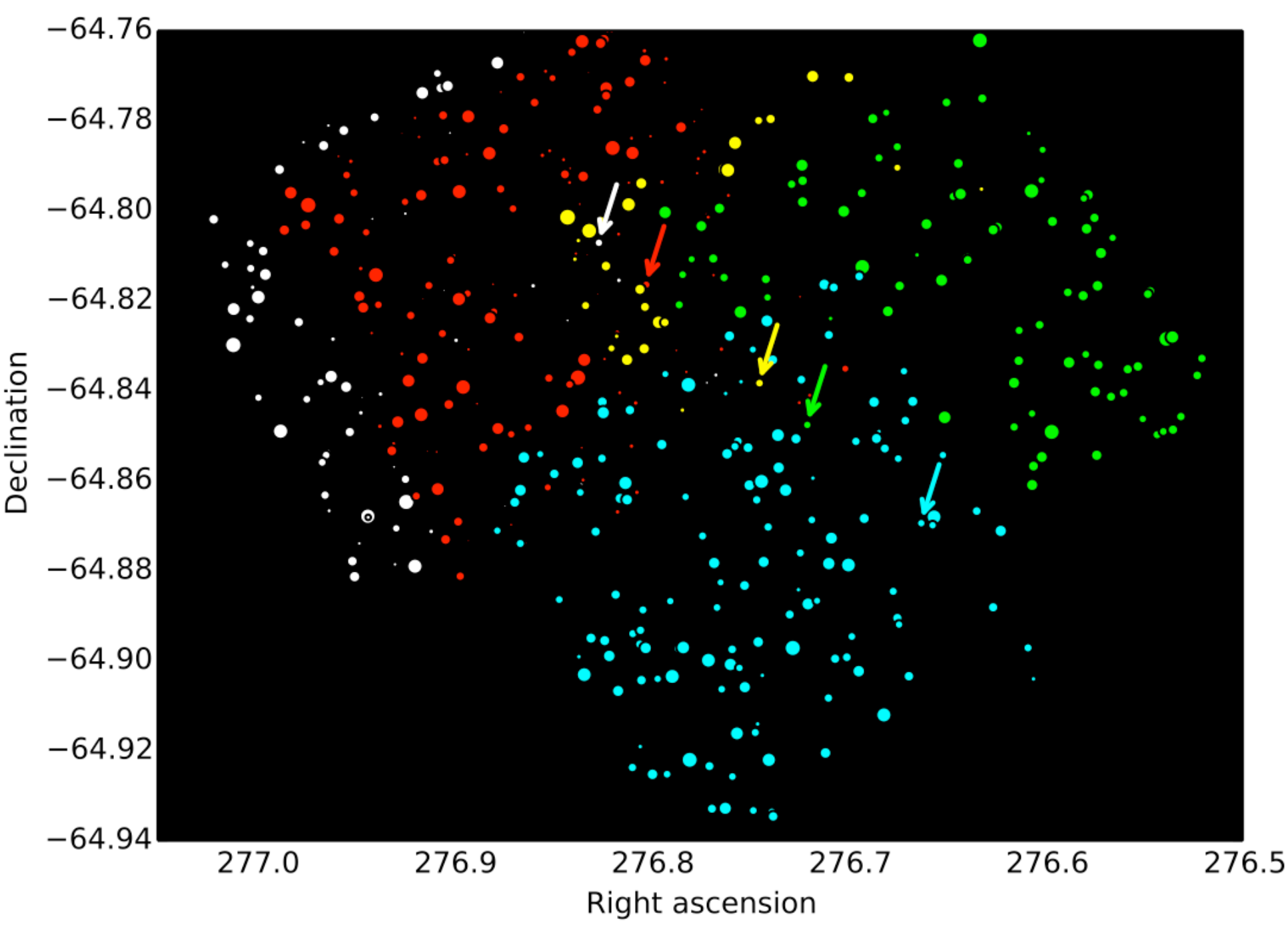}
		\caption{Each dot in this plot represents a source observed by Gaia, with the size representing different magnitudes. Five consecutive scans (each lasting 6 seconds) have been used to search for the asteroid (4997) Ksana, based on its predicted position. The arrows indicate the positions of the asteroid (4997) Ksana at each different scan, starting from the first (upper left) to the last one (lower right). The different colors are used to identify the different scans. Courtesy ESA/Gaia/DPAC/Airbus DS.	}
	\label{F:ksana}
\end{figure}

\section{Identification of asteroids in the data flow}
\label{S:ident}

The identification of Solar System objects in the data exploits the position and brightness of the source as reconstructed by the ''Initial Data Treatment'' (IDT) on the ground, which uses the daily attitude to perform a first, quick astrometric reduction with a very short delay from the observation. IDT considers data packets containing variable amounts of data, and for each packet performs the so-called cross-matching. This complex procedure identifies stars in the data packet by matching the position of the sources, to previously catalogued detections.

As the asteroid moves with a substantial displacement from one observation to the other, cross-matching fails to match its positions. Un-matched sources are provided to CU4 as candidate moving objects.

Within the SSO-ST pipeline, receiving the IDT output, a search algorithm (Berthier 2006) attempts to match the position and the (approximate) magnitude of each observed source to the ephemerides of each asteroid. The principles for computing the theoretical position of the asteroid are similar to those illustrated in Sect.~\ref{S:predict}, but the details are different, as more stringent requirements for the computation of speed are present. Also, the sources of some ancillary data required for the computation are different, as for instance the IDT data flow itself contains the position of Gaia at the epoch of each observation. Eventually, the equatorial coordinates of the observation computed by IDT are used, as they can be directly compared to the ephemerides, instead of predicting focal plane coordinates (a process that adds further uncertainties).

The source of orbital elements is the commonly used ASTORB database maintained at Lowell observatory (Flagstaff, AZ, USA).
As for un-numbered asteroids the discrepancy between prediction and observation can be rather high, a probabilistic approach is adopted to identify the most likely candidate for each source. In practice, when the ephemeris uncertainty contained in ASTORB (called CEU, ``Current Ephemeris Uncertainty") is of the order of a few arc-seconds or less, it is highly probable that only a single candidate asteroid can be associated to the prediction. Conversely, increasing orbital uncertainties result in higher and higher ambiguity of identification, as the object can fall into the overlapping uncertainty areas of several known sources. In such a situation the possible object identities are rated according to each orbital uncertainty and to the distance from the detected position. A ``probability of identification'' is assigned based on this criteria. The top element of the list, having the highest probability, is considered to be the most plausible candidate. If that match is wrong, the anomaly will be identified further downstream in the processing (for example it can exhibit a high residual when orbital fitting is attempted) and the alternative identities can be tested.

Failing the identification with respect to the data base of a moving object -- i.e. low probability of identification -- reveals the possible presence of a new asteroid and may result in the triggering of ground-based follow up. However, up to now (summer 2015), a high number of ``contaminants'' (i.e. unmatched sources that are not asteroids), due to the non perfect efficiency of IDT in the stellar cross-matching, prevents the use of SSO-ST for the original goal of alert triggering on unknown targets. A strategy to overcome this problem has been identified and is being tested. Only when the influence of contaminants will be negligible, SSO-ST will release the alerts to the community.

\section{Analysis of the asteroid signals}
\label{S:signal}

The processing of the astrometry starts in a first module (``CCD processing'') with the analysis of the CCD counts in order to determine the relevant parameters of the signals of the asteroids. In SSO-ST such parameters are the mean position (centroid) of the observed source for each CCD, and the associated flux. 

Centroid coordinates are raw quantities defined only in the space of the CCD window samples collected by the Gaia instrument. In short-term processing, centroids are computed by assuming a star-like PSF, that is by assuming, as a first approximation, that the source is not smeared by proper motion and not extended. While angular extension has an impact depending upon the size of the asteroid, smearing due to motion is nearly always present. In fact, the TDI mode of the CCD assumes that the source is moving across the field of view only due to the satellite rotation (i.e. as a fixed source, a star does). The asteroid proper motion relative to stars induces an image smearing. By fitting this signal with a model not taking motion into account, a deterioration of the astrometry is present. However, this approximation is fully acceptable and consistent with the accuracy requirements in short-term processing, as much higher uncertainty sources (in particular the attitude, see Sect.\ref{S:astrometry}) are present. 

As an example, we can consider a typical Main Belt, moving at $\sim$10 mas/s relative to stars. While crossing a single CCD (in 4.4 s) in the AF instrument, the image smearing can reach 44 mas, i.e. 73$\%$ of the pixel size. The error on the centroiding due to the assumption of a fixed source, will be a fraction of that quantity. However, the error on the attitude (Sect.\ref{S:astrometry}) will be much larger in the short-term processing (typically 100 mas) fully dominating the statistics (Sect.\ref{S:astrometry}).

On the other hand, despite the fact that centroids are accurate enough, we cannot expect that the {\it goodness of fit}, expressed in terms of the reduced $\chi^2$ of the difference between the observed signal and the PSF model, is the same we would have if the uncertainty was dominated only by mere photon noise statistics. 

\begin{figure}
	\centering
		\includegraphics[width=0.7\columnwidth]{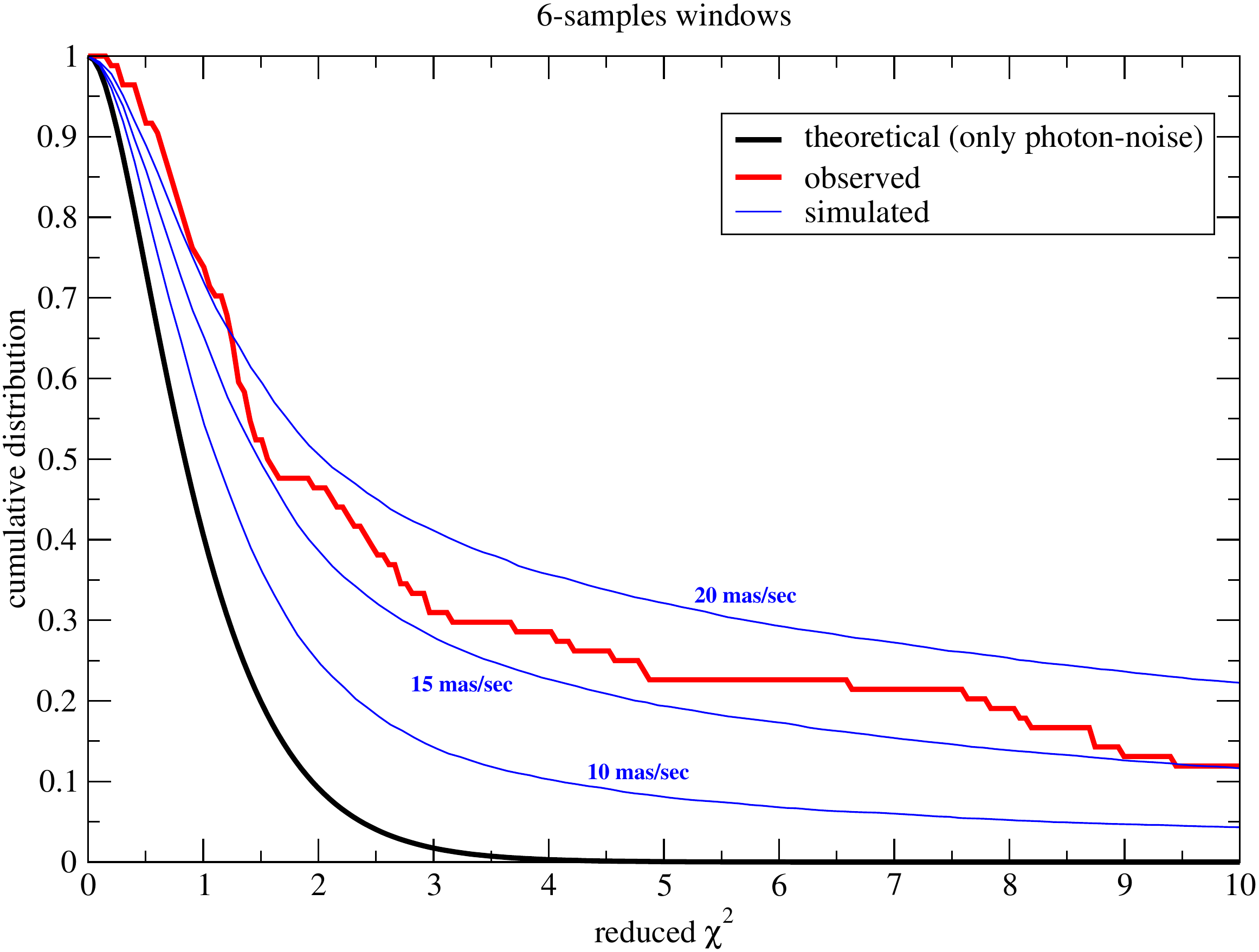}
		\includegraphics[width=0.7\columnwidth]{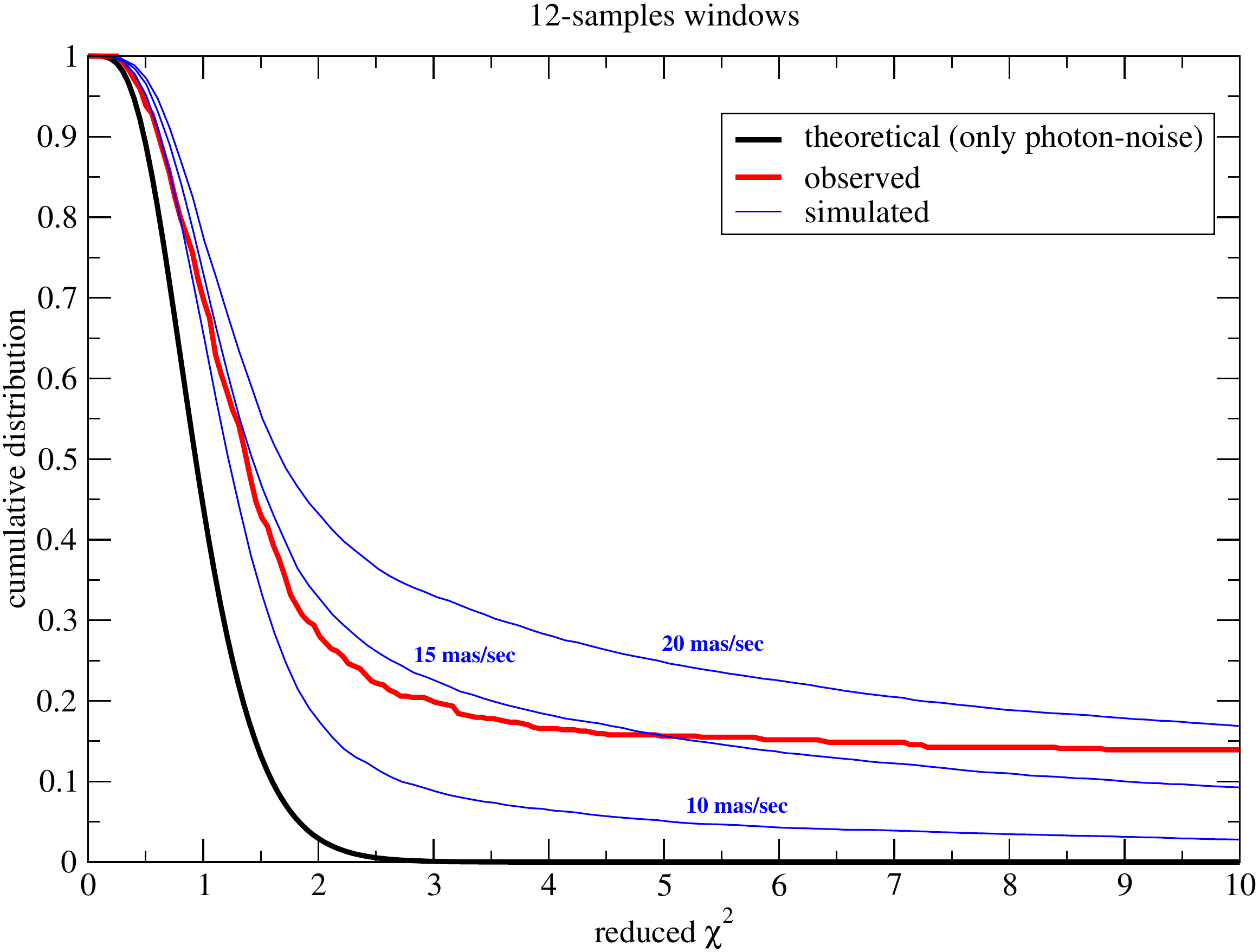}
		\caption{Cumulative distribution of the $\chi^2$, that is the fraction of centroids with $\chi^2$ larger than the values in abscissa. The red line is the distribution of the observed $\chi^2$, while the black line is the theoretical distribution of the signal with simulated photon noise included. Blue lines represent simulations with three different values of the proper motion dispersion (10, 15 and 20 mas/s). Top panel: 6-pixels window; bottom panel: 12-pixels window. The black line represents the theoretical curve in presence of photon noise.}
	\label{F:DOF}
\end{figure}

Fig.~\ref{F:DOF} shows the analysis of the distribution of the $\chi^2$ for a sample of $\sim$300 real asteroid signals provided for validation tests, for two window classes, corresponding to objects fainter than V$<$16 and for objects 13$<$V$<$16 ($6$-pixels and $12$-pixels width windows, respectively). The difference between the observed (red) and theoretical (black) distribution is mainly due to smearing by source motion.

A quality control of the data has been performed by simulating the distribution of the $\chi^2$. Our simulation takes into account both the distribution of the magnitudes of the observed sample and the distribution 
of the along-scan motion. The distribution of the along-scan motions of the asteroids is generated by a Gaussian distribution, with null average, reproducing rather well the simulated velocities. 

Simulations on long time intervals (5 years of nominal mission) exhibit a standard deviation of the along-scan motion of about $10$ mas/s for Main Belt asteroids, and $30$ mas/s for Near Earth Objects. 
Our analysis matches a velocity dispersion of about $15$ mas/s which might also be biased, with respect to predictions, by the narrow interval of time considered. As a consequence we naturally assume that 
the distribution of the velocities of such a subset cannot match exactly the expected distribution of the whole population and for the entire duration of the mission.

On the other hand, this result shows that the expectations on the centroiding accuracy are completely realistic and correspond to the real performance of Gaia.

\section{Astrometric data reduction}
\label{S:astrometry}

The positions of the asteroids derived from the signal acquired by Gaia are expressed in pixel coordinates on the CCDs. The transformation from pixel coordinates on each CCD to right ascension-declination pairs in the Barycentric Reference System (BCRS) is the task of an appropriate software module. In the same way the timings, given initially in OBMT, On-Board Mission Timeline (a technical timescale of Gaia) are converted to Barycentric Coordinate Time, TCB, the time-like coordinate of BCRS.

This coordinate transformation is heavily dependent on the framework set up by the core CUs of DPAC, and in particular on the results obtained by IDT (Initial Data Treatment), FL (First Look), ODAS (One-Day Astrometric Solution) and AGIS (Astrometric Global Iterative Solution), and on the software produced by the core CUs, made available to the Gaia community through an appropriate library.

Accurate transformations from CCD coordinates to sky coordinates, as computed by AGIS, are not available before many months after the observations, so in SSO-ST the low-precision ODAS solution is adopted to convert CCD coordinates to positions in the sky. Since in the course of one day, Gaia is mainly scanning along a great circle in the sky, reasonable accuracy in the transformation is available in the so-called along-scan direction, but in the perpendicular direction, across-scan, the transformation is less well constrained. 

On the average there will be a delay of one to two days between the epoch of observation, and its ground-based processing. This delay in producing an alert is critical for the recovery of the asteroids and all the efforts are spent to keep it below 3 days.

\begin{figure}
	\centering
		\includegraphics[width=0.7\columnwidth]{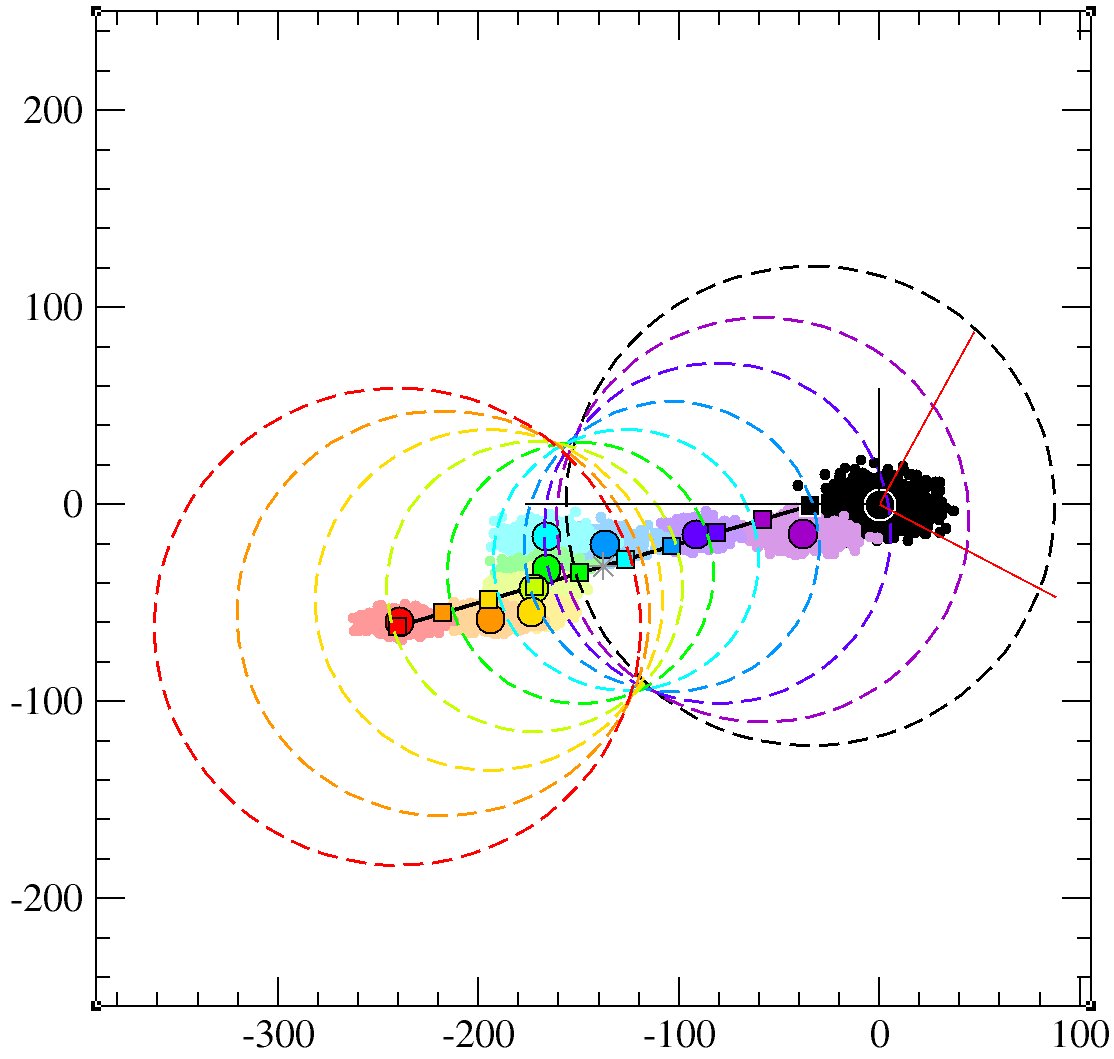}
		\caption{Validation plot for the analysis of a single transit. It is a representation of a small portion of the sky showing the different derived positions for a bright asteroid. The figure is oriented in such a way that the scan direction is vertical. The axis values are in mas. There are 10 positions, corresponding to the SM CCD (black, arbitrarily chosen as origin) and all 9 AF CCDs (rainbow colors from violet to red). The picture is supposed to show the motion in the sky of an SSO over $\sim$one minute (the duration of a transit). The colored circles (with solid contours) represent the measured positions converted to sky coordinates. Each of them is surrounded by a compact cloud of dots (lighter color) representing the scattering on the positions as derived from the centroiding uncertainty.  The colored squares represent a fit of a linear motion to the derived positions. The dashed circles represent an error budget containing the uncertainties on the fit plus the errors on the daily attitude. Rather than showing a linear motion, the derived positions (solid circles) show a path more resembling a corkscrew motion, which deviates significantly from a linear motion, if one considers the errors on the fit of the PSF to the images alone.  But if one considers also the errors on the attitude, the deviations are still well within the uncertainties.  It is not clear what the origin of the corkscrew motion is, but it may imply that there is a rotational residual in the attitude with a period of the order of 1 minute and an amplitude of the order of a few tens of mas.}
	\label{F:astrometry}
\end{figure}

Ancillary tasks of this software module are dedicated to select the appropriate positions for further processing. For instance, an asteroid may, due to its motion in the sky, leave the transmitted window before reaching the last CCD. In that case a spurious centroiding, not corresponding to the object, can be computed by the ``CCD processing''. As a consequence, the positions in the sky will no longer follow a linear motion, which should be the case in the course of one regular transit (about one minute).  By fitting a linear motion to the positions, it is possible to reject such outliers.

Also, by the same procedure, a so-called ``average'' transit position and a transit speed are derived. These transit positions and speeds are not intended for publication but can be used to link together observations, separated in time, of a same (unknown) asteroid. Orbits, however, will be computed from the individual positions for each CCD.

Such individual positions of the new asteroids will be sent to the Minor Planet Center. In SSO-ST this is done as quickly as possible, with short batches of a few positions. It should be noted that timings sent to the Minor Planet Center will not be in TCB (as will be in the output catalogue), but rather converted to the more user-friendly UTC.

Finally, an important task is the validation of the data. Both in the simulation phase and with real data, a number of checks are done to see whether the data produced corresponds to the expectations.  The most important check is that within a transit (about one minute) the different positions of an SSO show a linear motion in the sky within the quoted uncertainties. To this end, the software collects statistics of all objects processed, and generates user-defined plots to visualize these statistics, but also generates plots of single transits, to check whether a given SSO exhibits the expected linear motion or not. 

The example in Fig.~\ref{F:astrometry} shows a possible result from single-transit analysis. Not all the details shown in this plot are fully understood at present, such as the ``corkscrew motion'' around the average displacement, that could be partially related to the lower resolution (a factor 3) in the across-scan direction (horizontal axis) and to a preliminary geometric calibration of the focal plane. 

The final uncertainty on each position represented by the dashed circles (50 to 100 mas in radius) is consistent with the low accuracy of the daily attitude solution, independent from the object magnitude, that will collapse in the global astrometric solution. At the end of mission, the expected single-epoch accuracy for a V=20 asteroid with a ``slow'' proper motion (at the level of an average Main Belt) will be around 1.5 mas, while at V=15 it will approach 0.1 mas. 

As a consequence, there is a huge difference between the precision of the observations released on a daily basis, and the final ones. Incidentally, we can anticipate that asteroid positions reduced with accurate astrometric solutions will probably be released periodically in the planned Gaia Data Releases, and sent to MPC at the same time. Further details on the expected final accuracy can be found in Tanga and Mignard (2012). 

\section{Orbital inversion}
\label{S:inversion}

For all candidate ``new'' asteroids seen by Gaia, a short-arc initial orbit is required, for ground-based recovery.

Within Gaia DPAC CU4 object processing, initial orbital inversion is carried out for Solar System objects using random-walk statistical ranging (Muinonen {\it et al.} 2015), a newly developed method based on Markov chain Monte Carlo
(MCMC).  Random-walk ranging derives from a number of earlier ranging
methods (Virtanen {\it et al.}  2001, Muinonen {\it et al.} 2001) and the MCMC ranging method by
Oszkiewicz {\it et al.}  (2009). They start from the selection of a pair of astrometric observations, whereafter the gaiacentric ranges and angular deviations in Right Ascension (R.A.)
and Declination (Decl.)  are randomly sampled. Orbital elements then follow from the two Cartesian positions, obtaining probabilistic weights on the basis of the specific ranging method in question.

We describe the six osculating orbital elements of an asteroid at a given epoch $t_0$ by the vector ${\b P}$. For Cartesian elements,
${\b P} = (X,Y,Z,\dot{X},\dot{Y},\dot{Z})^T$, where, in a given
reference frame, the vectors $(X,Y,Z)^T$ and
$(\dot{X},\dot{Y},\dot{Z})^T$ denote the position and velocity,
respectively. Let $p_{\rm p}$ be the orbital-element probability
density function (p.d.f.). Within the Bayesian framework, $p_{\rm p}$
is proportional to the a priori and observational error p.d.f.s
$p_{\rm pr}$ and $p_{\epsilon}$, the latter being evaluated for the
sky-plane (``Observed-Computed'') residuals $\Delta{\b \psi}({\b P})$
(Muinonen and Bowell, 1993),
\begin{eqnarray}
p_{\rm p}({\b P}) &\propto& 
p_{\rm pr}({\b P}) p_{\epsilon}(\Delta {\b \psi}({\b P})),
\nonumber\\
\Delta{\b \psi}({\b P}) &=& {\b \psi} - {\b \Psi}({\b P}),
\label{eq:apost}
\end{eqnarray}
where ${\b \psi}$ and ${\b \Psi}$ denote the observations and the
computed positions.  $p_{\epsilon}$ is typically assumed to be
Gaussian.  The final a posteriori p.d.f. is then
\begin{eqnarray}
p_{\rm p}({\b P}) &\propto& \exp \left[ -\frac{1}{2} \chi^2({\b P}) \right],
\nonumber\\
\chi^2({\b P}) &=& 
\Delta{\b \psi}^T({\b P}) \Lambda^{-1} \Delta{\b \psi}({\b P}).
\label{eq:apost2}
\end{eqnarray}

The random-walk ranging method for sampling the a posteriori
probability density is implemented in the ''Short-arc orbit determination'' package DPAC CU4
software at CNES, Toulouse, France. The input consists of individual astrometric
positions for an object. The orbital computation results (i.e., 2000 sample orbits computed by
random-walk ranging) are passed through to the rest of the chain for ephemeris
prediction to be diffused to the ground-based follow-up network Gaia-FUN-SSO.

As an example to illustrate the performance of the approach, we compute the
orbital distribution of a single anonymous object observed by Gaia
around November 8, 2014. The data consist of 55 observations over 19 h 55.2 min.

\begin{figure}
	\centering
		\includegraphics[width=\columnwidth]{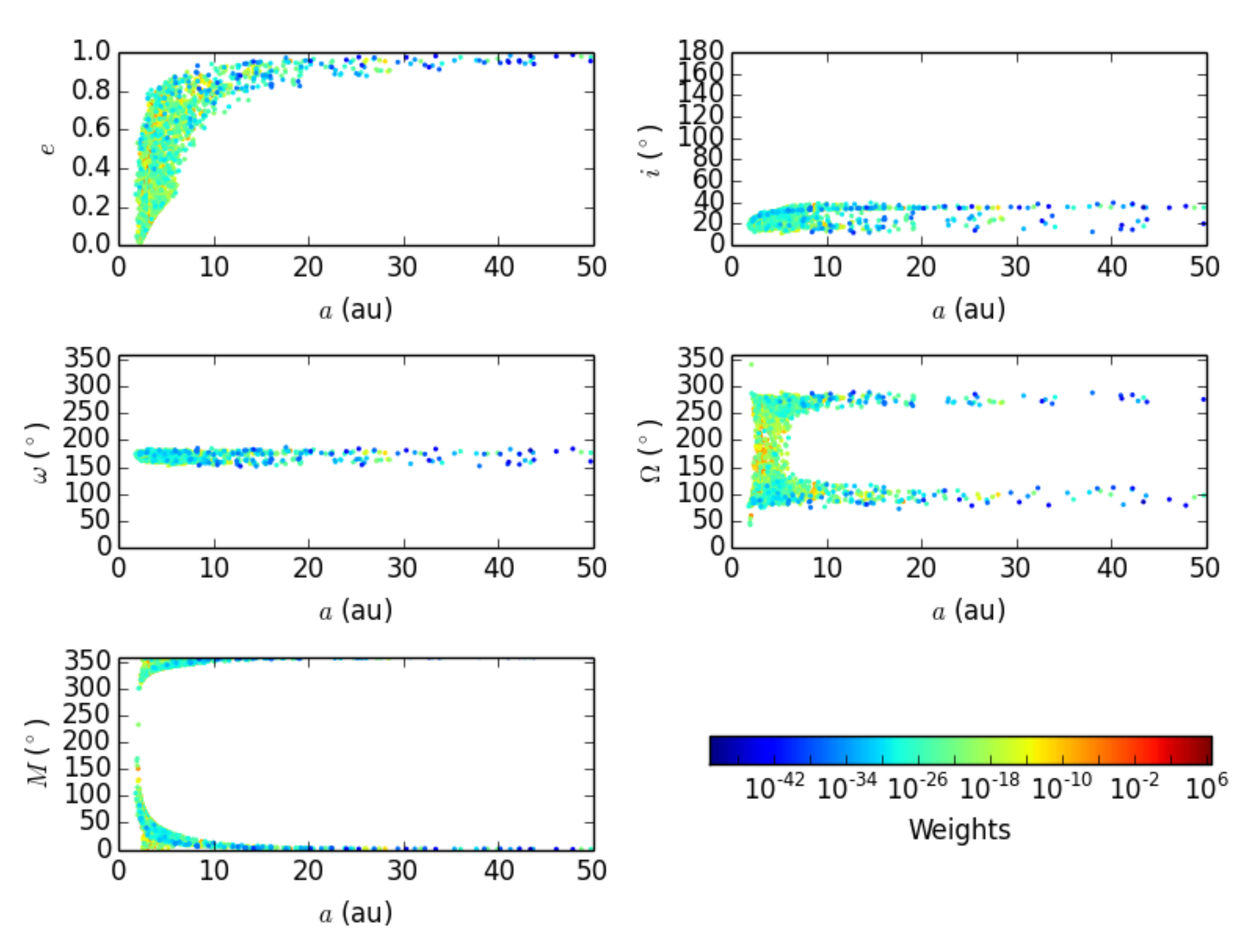}
		\caption{Keplerian orbital elements from random-walk
                  ranging for one of the objects with 55 Gaia observations
                  from two different transits. The asteroid is likely
                  to be a Main Belt object, and the weights already indicate a
                  preferred phase-space regime. }
	\label{F:orbit}
\end{figure}

The initial guess for the range is 2.5$\pm$ 0.2 AU using the
Gaussian probability distribution, based on the assumption that the
majority of the observed asteroids are main-belt objects. The
algorithm, however, also takes into consideration other possible
orbital types, by changing the initial guess for the range into
uniform sampling of ranges, should the initial guess fail during the
first 50 attempts. In Fig.~\ref{F:orbit} we illustrate the results using
Keplerian orbital elements.

We have also observed the so-called phase transition (e.g., Muinonen et al., 2006) to occur around the 12-h observational time interval. After the phase transition, corresponding to a substantially smaller region in the orbital parameter space, the ephemeris prediction is
constrained into a relatively small portion of the sky, which is a major aid for follow-up observations. However, the tentative conclusion has been reached on the basis of only several tens of observations, and therefore additional data are required for confirmation. The length of the observational time interval varies with different asteroid orbital type and may be used as an indirect means to
distinguish between different orbit types also in the Gaia data set.

\section{Ground-based follow up}
\label{S:followup}

The ground-based Follow-Up network for the Solar System Objects observed by Gaia (Gaia-FUN-SSO) is coordinated within DPAC, but relies on a network of that is completely external to the DPAC itself.

Nominally, it is designed to operate on newly discovered asteroids, but since SSO-ST is not yet operating in discovery mode, the network was mainly activated on test target, with training goals. 
Gaia-FUN-SSO is entirely managed at IMCCE, Paris, where a new web interface for registering the users, automatically disseminating the alerts and collecting the observations, was implemented (https://gaiafunsso.imcce.fr/).

The Gaia-FUN-SSO network has been set up on a volunteering base and gathers 56 observing sites equipped with 80 telescopes ensuring a good geographical coverage on Earth and some redundancy for overcoming bad meteorological conditions. This coverage, could be improved by expanding further to the southern hemisphere and North America. Nevertheless, such a big number of participants ensures a good potential when alerts are triggered.

Within the network almost 30 telescopes (those with diameter larger than 0.8m) are capable of tracking the Gaia discoveries close to the mission limit in brightness (V=20). 

\section{Conclusions}
\label{sec:concl}
We presented a quick overview of the main steps for Solar System processing of Gaia data. Despite the fact that these analysis are still very preliminary, we can say that the results obtained on validation data, extracted from the large volume of Gaia observations, appear to be consistent with the expectations. 
In particular, the single-epoch astrometric positions of asteroids on the sky are well within a dispersion of $\sim$70 mas, as allowed by the properties of the approximate, daily attitude. 
In future months Gaia observations will be used to trigger alerts on objects of special interest, and the long-term pipeline, implementing more accurate solutions, will also be tested, in the attempt to produce astrometry to be published in one of the first intermediate releases of Gaia.

\vspace*{5mm}

{\bf Acknowledgments.}  We gratefully acknowledge the Gaia team at CNES (Toulouse, France) where the Solar System data are processed. This research is funded, in part, by the Academy of
Finland (KM's project 1257966). GF acknowledges the support of the
Magnus Ehrnrooth foundation. ADO and AC acknowledge financial support from the Italian Space Agency (ASI), under the contracts `Gaia Mission - The Italian
Participation to DPAC'', 2014-025-R.0. TP is supported by Prodex. PT acknowledges support from CNES, and the Action Sp\'ecifique Gaia.

%% The Appendices part is started with the command \appendix;
%% appendix sections are then done as normal sections
%% \appendix

%% \section{}
%% \label{}

%% If you have bibdatabase file and want bibtex to generate the
%% bibitems, please use
%%
%%  \bibliographystyle{elsarticle-num} 
%%  \bibliography{<your bibdatabase>}

%% else use the following coding to input the bibitems directly in the
%% TeX file.

\end{document}